January 17, 2008

# Vector Matrices Realization of Hurwitz Algebras

Daniel Sepunaru

RCQCE - Research Center for Quantum Communication,

Holon Academic Institute of Technology,

52 Golomb St., Holon 58102, Israel

**Abstract**

We present the realization of Hurwitz algebras in terms of 2x2 vector matrices, which maintain the correspondence between the geometry of the vector spaces used in the classical physics and the underlined algebraic foundation of the quantum theory. The multiplication rule used is modification of the one originally introduced by M.Zorn. We demonstrate that our multiplication is not intrinsically non-associative; the realization of the real and complex numbers is commutative and associative, the real quaternions maintain associativity and the real octonion matrices form an alternative algebra. The extension to the calculus of the matrices (with Hurwitz algebra valued matrix elements) of the arbitrary dimensions is straightforward. We discuss briefly the applications of the obtained results to the extensions of the standard Hilbert space formulation of the quantum physics and to the alternative wave mechanical formulation of the classical field theory.

The mathematical formalism of the classical physics is based on the use of the real vector spaces. In contrast, the quantum physics typically formulated algebraically. It is desirable to work with structures that allow establishing the connection between both descriptions.

Among the diversity of possible algebras relevant in the physical applications, the Hurwitz algebras play a special role. They are the 1, 2, 4 and 8 dimensions quadratic normal division algebras and form the only possible numerical systems. We may generate using them the sequence of the mathematical frameworks suitable for the description of classical field theories (dispersion free [1]) as well as quantum field theories (that obey Heisenberg dispersion relations) using the Hilbert modules, functional analytical structures similar to the usual Hilbert spaces.

Consider subsequence of those structures with real scalar product, all dynamical variables mutually commuting and the states are real, complex, quaternion and octonion valued [2]:

$$(f,g)_R \equiv Tr(f,g) \tag{1}$$

where, for example, for $f$ and $g$ quaternion valued, $(f,g)$ is quaternion valued as well. The same structure may be generated alternatively by the four dimension vectors:

$$(f,g) = Tr(f,g) - e_1 Tr\{(f,g)e_1\} - e_2 Tr\{(f,g)e_2\} - e_3 Tr\{(f,g)e_3\} \tag{2}$$

$$-e_1(f,g)e_1 = Tr(f,g) - e_1 Tr\{(f,g)e_1\} + e_2 Tr\{(f,g)e_2\} + e_3 Tr\{(f,g)e_3\} \tag{3}$$

$$-e_2(f,g)e_2 = Tr(f,g) + e_1 Tr\{(f,g)e_1\} - e_2 Tr\{(f,g)e_2\} + e_3 Tr\{(f,g)e_3\} \tag{4}$$

$$-e_3(f,g)e_3 = Tr(f,g) + e_1 Tr\{(f,g)e_1\} + e_2 Tr\{(f,g)e_2\} - e_3 Tr\{(f,g)e_3\} \tag{5}$$

The sum of Eqs.(2), (3), (4) and (5) give us

$$(f,g)_R \equiv Tr(f,g) = \tfrac{1}{4}[(f,g) - e_1(f,g)e_1 - e_2(f,g)e_2 - e_3(f,g)e_3] \tag{6}$$

or in the matrix notations

$$(\Psi,\Phi)_R = \tfrac{1}{4}[\bar{f}, -e_1\bar{f}, -e_2\bar{f}, -e_3\bar{f}] * \begin{bmatrix} g \\ ge_1 \\ ge_2 \\ ge_3 \end{bmatrix} \tag{7}$$

Similarly, the Hilbert module with the complex scalar product is generated by the sum of Eq. (2) and Eq. (3):

$$(f,g)_C \equiv Tr(f,g) - e_1 Tr\{(f,g)e_1\} = (f,g) - e_1(f,g)e_1 \tag{8}$$

In the matrix notations

$$(\Psi, \Phi)_C = \tfrac{1}{2}[\bar{f}, -e_1 \bar{f})] * \begin{bmatrix} g \\ g e_1 \end{bmatrix} \tag{9}$$

The Hilbert module with the complex scalar product and the octonion valued states is generated exactly in the same manner. The usual Hilbert space obviously fits that procedure. This provides the evidence that the uniform matrix treatment of all Hurwitz algebras should exist.

First of all, let us consider 2x2 matrices. We have no difficulties to represent reals, complex and real quaternions but the underlined Cayley-Dickson procedure prevents the 2x2 matrix extension to the 8-dimensional algebra of real octonions. In addition, the obtained via Cayley-Dickson realization of the real quaternions

$$q = \begin{pmatrix} q_0 - i q_3 & -i q_1 - q_2 \\ -i q_1 + q_2 & q_0 + i q_3 \end{pmatrix} \tag{10}$$

provide the physically erroneous mapping of the space-time geometry

$$(x, y, z, t) \Rightarrow \begin{pmatrix} ct + z & x - iy \\ x + iy & ct - z \end{pmatrix} \tag{11}$$

since it violate the assumed isotropy of the space continuum. We therefore adopt the geometric vector matrices approach originally introduced by M.Zorn [3] [4] with the following modification:

1) Real numbers

$$X = x_0 \Rightarrow \begin{pmatrix} x_0 & 0 \\ 0 & x_0 \end{pmatrix}; \tag{12}$$

2) Complex numbers

$$X = x_0 + x_1 i \equiv x_0 + \vec{x} \Rightarrow \begin{pmatrix} x_0 & \vec{x} \\ \vec{x} & x_0 \end{pmatrix} = \begin{pmatrix} x_0 & x_1 i \\ x_1 i & x_0 \end{pmatrix} \tag{13}$$

3) Quaternions

$$X = x_0 + \sum_{i=1}^{3} x_i e_i \equiv x_0 + \vec{x} \Rightarrow \begin{pmatrix} x_0 & \vec{x} \\ \vec{x} & x_0 \end{pmatrix} = \begin{pmatrix} x_0 & \sum_{i=1}^{3} x_i e_i \\ \sum_{i=1}^{3} x_i e_i & x_0 \end{pmatrix} \qquad (14)$$

4) Octonions

$$X = x_0 + \sum_{i=1}^{7} x_i e_i \equiv x_0 + \vec{x} \Rightarrow \begin{pmatrix} x_0 & \vec{x} \\ \vec{x} & x_0 \end{pmatrix} = \begin{pmatrix} x_0 & \sum_{i=1}^{7} x_i e_i \\ \sum_{i=1}^{7} x_i e_i & x_0 \end{pmatrix} \qquad (15)$$

and the multiplication rule defined by

$$Z = X \Diamond Y \equiv \begin{pmatrix} x_0 & \vec{x} \\ \vec{x} & x_0 \end{pmatrix} \Diamond \begin{pmatrix} y_0 & \vec{y} \\ \vec{y} & y_0 \end{pmatrix} = \begin{pmatrix} x_0 y_0 + \vec{x} \bullet \vec{y} & x_0 \vec{y} + y_0 \vec{x} + \vec{x} \times \vec{y} \\ x_0 \vec{y} + y_0 \vec{x} + \vec{x} \times \vec{y} & x_0 y_0 + \vec{x} \bullet \vec{y} \end{pmatrix} \qquad (16)$$

where

$$e_i \bullet e_j = -\delta_{ij}$$

$$\vec{x} \bullet \vec{y} = -x_i y_i = \vec{y} \bullet \vec{x} \qquad (17)$$

$$\vec{x} \times \vec{y} = \varepsilon_{ijk} x_i y_j e_k = -\vec{y} \times \vec{x} \ ;$$

$\varepsilon_{ijk}$ are the structure constants of the correspondent multiplication table (see Appendix). For the quaternions it is usual totally antisymmetric three dimension tensor; in case of octonions it also may be considered as Levi-Chevita tensor in seven dimension space.

Explicitly, for the quaternions we have

$$\vec{x} \times \vec{y} = (x_2 y_3 - x_3 y_2) e_1 + (x_3 y_1 - x_1 y_3) e_2 + (x_1 y_2 - x_2 y_1) e_3 \qquad (18)$$

and for the octonions

$$\vec{x} \times \vec{y} = (x_2 y_3 - x_3 y_2) e_1 + (x_3 y_1 - x_1 y_3) e_2 + (x_1 y_2 - x_2 y_1) e_3$$

$$+ (x_6 y_5 - x_5 y_6) e_1 + (x_6 y_2 - x_2 y_6) e_4 + (x_2 y_5 - x_5 y_2) e_7$$

$$+ (x_4 y_7 - x_7 y_4) e_1 + (x_7 y_2 - x_2 y_7) e_5 + (x_2 y_4 - x_4 y_2) e_6$$

$$+ (x_4 y_6 - x_6 y_4)e_2 + (x_1 y_4 - x_4 y_1)e_7 + (x_1 y_6 - x_6 y_1)e_5 \qquad (19)$$

$$+ (x_5 y_7 - x_7 y_5)e_2 + (x_5 y_1 - x_1 y_5)e_6 + (x_7 y_1 - x_1 y_7)e_4$$

$$+ (x_6 y_7 - x_7 y_6)e_3 + (x_7 y_3 - x_3 y_7)e_6 + (x_3 y_6 - x_6 y_3)e_7$$

$$+ (x_5 y_4 - x_4 y_5)e_3 + (x_3 y_5 - x_5 y_3)e_4 + (x_4 y_3 - x_3 y_4)e_5$$

Obviously,

$$\vec{x} \times \vec{x} = 0 \quad \text{and} \quad \vec{x} \bullet \vec{x} = -\sum_i x_i^2 \qquad (20)$$

An involution is defined by

$$\overline{X} \equiv x_0 - \vec{x} \Rightarrow \begin{pmatrix} x_0 & -\vec{x} \\ -\vec{x} & x_0 \end{pmatrix} \qquad (21)$$

And satisfy the standard requirements

$$\overline{\overline{X}} = X$$

(follows immediately from (21))

$$\overline{X \Diamond Y} = \overline{Y} \Diamond \overline{X} \qquad (22)$$

Proof:

$$\overline{X \Diamond Y} = \begin{pmatrix} x_0 y_0 + \vec{x} \bullet \vec{y} & -x_0 \vec{y} - y_0 \vec{x} - \vec{x} \times \vec{y} \\ -x_0 \vec{y} - y_0 \vec{x} - \vec{x} \times \vec{y} & x_0 y_0 + \vec{x} \bullet \vec{y} \end{pmatrix} \qquad (23)$$

$$\overline{Y} \Diamond \overline{X} = \begin{pmatrix} y_0 & -\vec{y} \\ -\vec{y} & y_0 \end{pmatrix} \Diamond \begin{pmatrix} x_0 & -\vec{x} \\ -\vec{x} & x_0 \end{pmatrix} = \begin{pmatrix} x_0 y_0 + \vec{x} \bullet \vec{y} & -x_0 \vec{y} - y_0 \vec{x} - \vec{x} \times \vec{y} \\ -x_0 \vec{y} - y_0 \vec{x} - \vec{x} \times \vec{y} & x_0 y_0 + \vec{x} \bullet \vec{y} \end{pmatrix} \qquad (24)$$

Now we are in the position to proof the following statement: the algebras defined by Eqs. (12), (13), (14), (15), (16) and (17) are quadratic normal division algebras.

Proof:

1) $Tr(X) \equiv X + \overline{X} = 2x_0$

(25)

2) $N(X) \equiv X \Diamond \overline{X} = \begin{pmatrix} x_0 & \vec{x} \\ \vec{x} & x_0 \end{pmatrix} \Diamond \begin{pmatrix} x_0 & -\vec{x} \\ -\vec{x} & x_0 \end{pmatrix} = \begin{pmatrix} x_0^2 - \vec{x} \bullet \vec{x} & 0 \\ 0 & x_0^2 - \vec{x} \bullet \vec{x} \end{pmatrix}$

Then

$$X^2 - Tr(X)X + N(X) = \begin{pmatrix} x_0 & \vec{x} \\ \vec{x} & x_0 \end{pmatrix} \Diamond \begin{pmatrix} x_0 & -\vec{x} \\ -\vec{x} & x_0 \end{pmatrix} - 2x_0 \begin{pmatrix} x_0 & \vec{x} \\ \vec{x} & x_0 \end{pmatrix} + (x_0^2 - \vec{x} \bullet \vec{x})I = 0 \quad (26)$$

From the uniqueness of the Hurwitz algebras it follows that the discussed realization has the following properties:

1) One dimension algebra of reals and two dimension algebra of complex numbers

$$X \Diamond Y = Y \Diamond X \quad \text{(commutative)}$$

$$X \Diamond (Y \Diamond Z) = (X \Diamond Y) \Diamond Z \quad \text{(associative)} \quad (27)$$

2) Four dimension algebra of real quaternions

$$X \Diamond (Y \Diamond Z) = (X \Diamond Y) \Diamond Z \quad \text{(associative)} \quad (28)$$

3) Eight dimension algebra of real octonions

$$X^2 \Diamond Y = X \Diamond (X \Diamond Y) \quad \text{(left alternative)} \quad (29)$$

$$X \Diamond Y^2 = (X \Diamond Y) \Diamond Y \quad \text{(right alternative)} \quad (30)$$

Indeed, the validity the above statements may be demonstrated through direct matrix calculations. However, they are rather cumbrous and we will only provide the useful relations for that:

1) For all Hurwitz algebras hold

$$\vec{x} \bullet \vec{y} = \vec{y} \bullet \vec{x} \quad (31)$$

$$\vec{x} \times \vec{y} = -\vec{y} \times \vec{x} \quad (32)$$

$$\vec{x} \bullet (\vec{y} \times \vec{z}) = \vec{z} \bullet (\vec{x} \times \vec{y}) = \vec{y} \bullet (\vec{z} \times \vec{x}) \quad (33)$$

2) For Quaternions

$$\vec{x} \times (\vec{y} \times \vec{z}) = (\vec{x} \bullet \vec{y})\vec{z} - (\vec{x} \bullet \vec{z})\vec{y} \quad (34)$$

Using relations (33) and (34) we have

$$\begin{aligned}(x \Diamond y) \Diamond z &- x \Diamond (y \Diamond z) = \\ &= [(\vec{x} \times \vec{y}) \bullet \vec{z} - (\vec{y} \times \vec{z}) \bullet \vec{x}] + [(\vec{x} \bullet \vec{y})\vec{z} - (\vec{y} \bullet \vec{z})\vec{x} + (\vec{x} \times \vec{y}) \times \vec{z} - \vec{x} \times (\vec{y} \times \vec{z})] = 0\end{aligned} \quad (35)$$

3) For Octonions

$$\vec{x} \times (\vec{x} \times \vec{y}) = -(\vec{x} \bullet \vec{y})\vec{x} + (\vec{x} \bullet \vec{x})\vec{y} \tag{36}$$

Using (33) for the scalar component of the alternator we have

$$(\vec{x} \times \vec{y}) \bullet \vec{z} - (\vec{y} \times \vec{z}) \bullet \vec{x} = 0 \tag{37}$$

Therefore

$$Tr[(x \lozenge y) \lozenge z] = Tr[x \lozenge (y \lozenge z)] \tag{38}$$

Thus the calculations of the scalar products in the real Hilbert module with the octonion valued states may be performed neglecting their non-associativity. Obviously we have also

$$Tr[(x \lozenge y) \lozenge z] = Tr[z \lozenge (x \lozenge y)] \tag{39}$$

We have obtained the properties (associativity and commutativity) both needed to formulate dispersion free field theory [2].

The detailed discussion of self-adjoint operators (dynamical variables) in those frameworks will be presented in a separate publication.

Using (36) for the vector component of the alternator we have

$$\begin{aligned}(\vec{x} \bullet \vec{x})\vec{y} - (\vec{x} \bullet \vec{y})\vec{x} - \vec{x} \times (\vec{x} \times \vec{y}) = \\ = (\vec{x} \bullet \vec{x})\vec{y} - (\vec{x} \bullet \vec{y})\vec{x} + (\vec{x} \bullet \vec{y})\vec{x} - (\vec{x} \bullet \vec{x})\vec{y} = 0\end{aligned} \tag{40}$$

or

$$x^2 \lozenge y = x \lozenge (x \lozenge y) \qquad \text{(left alternative)} \tag{41}$$

Similarly,

$$\begin{aligned}(\vec{x} \bullet \vec{y})\vec{y} - (\vec{y} \bullet \vec{y})\vec{x} + (\vec{x} \times \vec{y}) \times \vec{y} = \\ = (\vec{x} \bullet \vec{y})\vec{y} - (\vec{y} \bullet \vec{y})\vec{x} - (\vec{x} \bullet \vec{y})\vec{y} + (\vec{y} \bullet \vec{y})\vec{x} = 0\end{aligned} \tag{42}$$

or

$$y \lozenge x^2 = (y \lozenge x) \lozenge x \qquad \text{(right alternative)} \tag{43}$$

Then the flexibility and the Moufang identities follow

$$(x \Diamond y) \Diamond x = x \Diamond (y \Diamond x) \tag{44}$$

$$(x \Diamond a \Diamond x) \Diamond y = x \Diamond [a \Diamond (x \Diamond y)] \tag{45}$$

$$y \Diamond (x \Diamond a \Diamond x) = [(y \Diamond x) \Diamond a] \Diamond x \tag{46}$$

$$(x \Diamond y) \Diamond (a \Diamond x) = x \Diamond (y \Diamond a) \Diamond x \tag{47}$$

Consider now the matrices of arbitrary dimension with matrix elements belong to one of the Hurwitz algebras. Then the product matrix is defined by the usual multiplication rule:

$$\begin{pmatrix} Z_{11} & Z_{12} & \cdots & Z_{1n} \\ Z_{21} & Z_{22} & \cdots & Z_{2n} \\ \vdots & \vdots & & \vdots \\ Z_{n1} & Z_{n2} & \cdots & Z_{nn} \end{pmatrix} \equiv \begin{pmatrix} X_{11} & X_{12} & \cdots & X_{1n} \\ X_{21} & X_{22} & \cdots & X_{2n} \\ \vdots & \vdots & & \vdots \\ X_{n1} & X_{n2} & \cdots & X_{nn} \end{pmatrix} \begin{pmatrix} Y_{11} & Y_{12} & \cdots & Y_{1n} \\ Y_{21} & Y_{22} & \cdots & Y_{2n} \\ \vdots & \vdots & & \vdots \\ Y_{n1} & Y_{n2} & \cdots & Y_{nn} \end{pmatrix} \tag{48}$$

$$Z_{ij} \equiv \sum_{k=1}^{n} X_{ik} Y_{kj} \;\; ; \;\; i,j = 1,2,...n \tag{49}$$

where

$$Z_{ij} \equiv \begin{pmatrix} Z_{ij}^0 & \vec{Z}_{ij} \\ \vec{Z}_{ij} & Z_{ij}^0 \end{pmatrix} = \sum_{k=1}^{n} \begin{pmatrix} x_{ik}^0 & \vec{x}_{ik} \\ \vec{x}_{ik} & x_{ik}^0 \end{pmatrix} \Diamond \begin{pmatrix} y_{kj}^0 & \vec{y}_{kj} \\ \vec{y}_{kj} & y_{kj}^0 \end{pmatrix} \equiv$$

$$\equiv \sum_{k=1}^{n} \begin{pmatrix} x_{ik}^0 y_{kj}^0 + \vec{x}_{ik} \bullet \vec{y}_{kj} & x_{ik}^0 \vec{y}_{kj} + y_{kj}^0 \vec{x}_{ik} + \vec{x}_{ik} \times \vec{y}_{kj} \\ x_{ik}^0 \vec{y}_{kj} + y_{kj}^0 \vec{x}_{ik} + \vec{x}_{ik} \times \vec{y}_{kj} & x_{ik}^0 y_{kj}^0 + \vec{x}_{ik} \bullet \vec{y}_{kj} \end{pmatrix} \tag{50}$$

$\forall X_{ik}, Y_{ik}, Z_{ik}$ elements of R,C,H and O algebras.

Thus the product matrix is defined as the usual sum of pairs of the multipliers and each pair product is defined by the vector multiplication introduced above.

In a conclusion, we have discussed the geometrical extension of the conventional matrix multiplication which uniformly valid for all quadratic normal division algebras. I would like to emphasize that the suggested matrix realization is of crucial importance for the quaternion and octonion extensions of the standard functional analysis since the real as well as complex Hilbert modules require the use of the multicomponent states. The obtained results allow introduction and investigation of the operators necessary for the description system dynamics and system dynamical observables (self-adjoint operators) [5] [6]. In addition, the transition from the vector matrices realization to the standard one may provide an alternative mechanism for the spontaneous breakdown of the internal symmetries as suggested the comparison between Eqs. (10) and (14). Historically, the multiplication operation over real numbers was first extended to the physically relevant three dimension space and only later to the spaces of the arbitrary dimension and signature [7]. The alternative direction of the generalization was the invention of the scalar matrix multiplication. It seems reasonable to expect that the suggested vector matrix multiplication may be extended to include the additional types of algebras (Clifford, Lie, Jordan, etc.) but that question lies outside the scope of my investigation.

I am grateful to L. Sepunaru for discussions.

**Appendix**

For the reader that is not lazy to verify my statements by the direct calculations, I reproduce here the Hurwitz algebras multiplication tables.

Table I

Complex numbers

|       | $e_0$  | $e_1$  |
|-------|--------|--------|
| $e_0$ | $e_0$  | $e_1$  |
| $e_1$ | $e_1$  | $-e_0$ |

Table II

Quaternions

|       | $e_0$ | $e_1$  | $e_2$  | $e_3$  |
|-------|-------|--------|--------|--------|
| $e_0$ | $e_0$ | $e_1$  | $e_2$  | $e_3$  |
| $e_1$ | $e_1$ | $-e_0$ | $e_3$  | $-e_2$ |
| $e_2$ | $e_2$ | $-e_3$ | $-e_0$ | $e_1$  |
| $e_3$ | $e_3$ | $e_2$  | $-e_1$ | $-e_0$ |

Table III

Octonions

|       | $e_0$ | $e_1$  | $e_2$  | $e_3$  | $e_4$  | $e_5$  | $e_6$  | $e_7$  |
|-------|-------|--------|--------|--------|--------|--------|--------|--------|
| $e_0$ | $e_0$ | $e_1$  | $e_2$  | $e_3$  | $e_4$  | $e_5$  | $e_6$  | $e_7$  |
| $e_1$ | $e_1$ | $-e_0$ | $e_3$  | $-e_2$ | $e_7$  | $-e_6$ | $e_5$  | $-e_4$ |
| $e_2$ | $e_2$ | $-e_3$ | $-e_0$ | $e_1$  | $e_6$  | $e_7$  | $-e_4$ | $-e_5$ |
| $e_3$ | $e_3$ | $e_2$  | $-e_1$ | $-e_0$ | $-e_5$ | $e_4$  | $e_7$  | $-e_6$ |
| $e_4$ | $e_4$ | $-e_7$ | $-e_6$ | $e_5$  | $-e_0$ | $-e_3$ | $e_2$  | $e_1$  |
| $e_5$ | $e_5$ | $e_6$  | $-e_7$ | $-e_4$ | $e_3$  | $-e_0$ | $-e_1$ | $e_2$  |
| $e_6$ | $e_6$ | $-e_5$ | $e_4$  | $-e_7$ | $-e_2$ | $e_1$  | $-e_0$ | $e_3$  |
| $e_7$ | $e_7$ | $e_4$  | $e_5$  | $e_6$  | $-e_1$ | $-e_2$ | $-e_3$ | $-e_0$ |